# Title: Investigation of multiple-dynein transport of melanosomes by non-invasive force measurement using fluctuation unit $\chi$

Shin Hasegawa[1], Takashi Sagawa[2], Kazuho Ikeda[3], Yasushi Okada[3,4,5], and Kumiko Hayashi[1,*]

[1]Department of Applied Physics, Graduate School of Engineering, Tohoku University, Sendai, Japan

[2]Advanced ICT Research Institute, National Institute of Information and Communications Technology, Kobe, Japan

[3]Laboratory for Cell Dynamics Observation, Center for Biosystems Dynamics Research, RIKEN, Osaka, Japan

[4]Department of Physics and Universal Biology Institute, Graduate School of Science, The University of Tokyo, Tokyo, Japan

[5]Department of Physics, Universal Biology Institute, and the International Research Center for Neurointelligence (WPI-IRCN), The University of Tokyo, Tokyo, Japan

*Correspondence: kumiko@camp.apph.tohoku.ac.jp (K.H.)

Pigment organelles known as melanosomes disperse or aggregate in a melanophore in response to hormones. These movements are mediated by the microtubule motors kinesin-2 and cytoplasmic dynein. However, the force generation mechanism of dynein, unlike that of kinesin, is not well understood. In this study, to address this issue, we investigated the dynein-mediated aggregation of melanosomes in zebrafish melanophores. We applied the fluctuation theorem of non-equilibrium statistical mechanics to estimate forces acting on melanosomes during transport by dynein, given that the energy of a system is related to its fluctuation. Our results demonstrate that multiple force-producing units cooperatively transport a single melanosome. Since the force is generated by dynein, this suggests that multiple dyneins carry a single melanosome. Cooperative transport has been reported for other organelles; thus, multiple-motor transport may be a universal mechanism for moving organelles within the cell.

## Introduction

Cellular cargo is transported through the microtubule network of eukaryotic cells by motor proteins[1,2]. This active transport system delivers materials more rapidly than the passive transport by diffusion occurring in prokaryotic cells. The importance of cargo transport is underscored by the fact that defects in this process in neurons are associated with neuronal diseases such as Alzheimer's, Parkinson's, and Huntington's disease[3,4].

Multiple-carrier (motor) transport of cargo has recently been proposed[5-13] as a mechanism for maintaining the stability of the intracellular transport system by increasing physical quantities such as force and run length. Since the cytosol has a higher viscosity than water[14-16], a significant amount of force must be generated through cooperation of multiple motors to overcome friction and haul cargo in cells.

In order to investigate force generated by motors acting on a single cargo, a non-invasive force measurement[11-13] was recently developed based on the fluctuation theorem of non-equilibrium statistical mechanics[17-22]. Fluctuation in small systems is not merely due to random motion, but is related to system energetics through physical theorems. The position of a moving cargo can be obtained non-invasively by fluorescence microscopic observation of cells, and the fluctuation of its position is easily observed with high time resolution. Using this fluctuating motion, a fluctuation unit ($\chi$) constructed based on the fluctuation theorem is calculated as a force indicator[11-13]. The physical background of $\chi$ was explained and discussed in the reference[13].

The parameter $\chi$ was previously measured for the transport of synaptic vesicle precursors (SVPs) by the kinesin superfamily protein UNC-104 in the DA9 neuron of *Caenorhabditis elegans*[12]. In wild-type worms, the distribution of $\chi$ was spread over several clusters, implying the existence of several force-producing units (FPUs); this in turn indicates that a single SVP was carried by multiple UNC-104 motors, which are the generators of force. Measurements of $\chi$ revealed that mutant worms lacking ARL-8—an SVP-bound Arf-like small guanosine triphosphatase that relieves the autoinhibition of the motor, which is critical for avoiding unnecessary consumption of ATP when the motor is not bound to an SVP[23]—had fewer FPUs than their wild-type counterparts[12].

In this study, we investigated $\chi$ for the transport of melanosomes, organelles filled with the melanin pigment, in zebrafish melanophores. Melanosomes are transported by microtubule motors, kinesin-2 and cytoplasmic dynein, and the actin motor myosin-V[24]. They disperse or aggregate in response to hormones in melanophores. The primary physiological purpose for movement of melanosomes in animals is colour change. Particularly, we focused on the aggregation process of melanosomes transported by dynein. There are several advantages of measurement of $\chi$ in this transport system. Firstly, because melanin pigment is black and easily observed by bright-field microscopy, the recording rate can be increased up to 800 frames per second (fps). Such a high-speed recording enables accurate measurement of fluctuation in the position of melanosomes. Secondly, we can decrease the number of dynein motors using the inhibitor ciliobrevin, which was recently identified[25] and investigated for many biological phenomena related to dynein[25-29], thereby allowing observation of the behaviour of fluctuation unit $\chi$ in response to the decrease in motors. Thirdly, the force generation mechanism of dynein is still controversial and therefore worth studying. The stall force value for a single dynein molecule was reported to be 1 pN[9,10,30], though a few studies have reported values of 5–10 pN[31-33]. Fourthly, the behaviour of $\chi$ remains to be evaluated in non-neuronal cells. When considered together, melanosome transport is a suitable system to explore $\chi$.

In our experiments, the quantal behaviour of $\chi$ was observed in melanosome transport by dynein. We concluded that several FPUs carry a single melanosome similarly to cargo transport in neurons[11-13]. Further, we found that the number of FPUs measured by $\chi$ was dramatically decreased by the addition of ciliobrevin, as expected. We anticipate that the non-invasive force measurement using the fluctuation unit ($\chi$) will be applied to a wide range of cargo transport in eukaryotic cells to elucidate the physical mechanisms in which some human diseases are deeply rooted.

**Results**

*Observation of melanosome transport*

Melanophores removed from zebrafish scales by enzymatic treatment were cultured in a glass-bottom dish. After 1 day of culture, the motion of melanosomes was observed by bright-field microscopy. Following addition of hormone (epinephrine) to the dish, melanosomes aggregated to the centre of a melanophore. Because the minus ends of microtubules are attached to the centrosome, the motion towards the nucleus is caused by the action of dynein. A schematic of this transport is represented in Fig. 1.

Although they were crowded at the start of migration, individual melanosomes could be tracked after several tens of seconds, when nearly all melanosomes were aggregated. The direction of movement was set as the plus $X$ direction (e.g. red line in Fig. 2a). Except for melanosomes close to the nucleus where cells had greater thickness, we considered that the motion in the $Z$ direction was less than that in the $X$-$Y$ plane since the images of melanosomes were focused during the runs. The centre position ($X$) of the melanosome was calculated from the recorded images, and the time course of $X$ was obtained (Fig. 2b). Directional movement was defined as movement at a velocity greater than 100 nm/s, as in a previous study on pigment granule transport[5]. The fluctuation in melanosome position during directional transport, which was observed at a high recording rate of 800 fps (inset in Fig. 2b), was mainly due to thermal noise, stochastic stepping of motors accompanied by ATP hydrolysis, and collision of melanosomes with other organelles and cytoskeletons.

*Calculation of $\chi$ for melanin pigment transport*

The constant velocity segment (red area in Fig. 2b) of the position ($X$) of a melanosome lasting 0.5–1 s was identified by visual inspection for calculation of the fluctuation unit $\chi$ (equation (4)). The effect of this selection on the calculation of $\chi$ was later evaluated by the bootstrapping method. Based on this segment, we calculated $\Delta X = X(t + \Delta t) - X(t)$ (e.g. inset in Fig. 2b) and its probability distribution ($P(\Delta X)$) (Fig. 2c for the case $\Delta t = 37.5$ ms), where the value of $\Delta t$ ranged from 1.25 to 37.5 ms. $P(\Delta X)$ was well fitted by a Gaussian function (solid black curve in Fig. 2c). Because $\Delta X < 0$ could not be observed for a large $\Delta t$ because of the limitations of our experiment, $P(\Delta X)$ for $\Delta X > 0$ was fitted with a Gaussian function to calculate the ratio $P(\Delta X)/P(-\Delta X)$ in equation (4). The Gaussian function (equation (5)) was used to fit $P(\Delta X)$ after evaluating the Brownian noise power spectrum density $S(f)$ (equation (11)) of $X$ for small $f$ (Supplementary Fig. S1).

$\chi$ defined in equation (4) was calculated using equation (6) for each $\Delta t$ (Fig. 2d); $\chi$ as a

function of $\Delta t$ converged to a constant value ($\chi^*$) after relaxation ($\sim 20$ ms), which is attributable to the microscopic environment around the melanosome as well as the enzymatic cycle time[13]. Note that from a physical point of view, the convergence implied that the macroscopic motion of $X$ whose time scale is about 20 ms obeys the simple equation (3). Then, equation (7) was considered for the constant value $\chi^*$.

The error of $\chi^*$ according to range selection of the constant velocity segment was estimated to be 15% based on the principle of bootstrapping. We first selected 10 different partial segments from the original constant velocity segments; $\chi$ was then calculated for each partial segment (thin curves in Fig. 2d). It should be noted that the length of partial segments was half that of the original segment.

*Melanosome transport by multiple dynein motors*

The procedures carried out in Fig. 2b–d for a single melanosome were repeated for 62 different melanosomes from two different melanophore preparations. The calculated values of $\chi$ for the 62 melanosomes are shown in Fig. 3a. These data were classified into four groups by applying affinity propagation (cluster) analysis (Fig. 3b). When we examined the $q$ value—the sole parameter of this analysis—a cluster number of 4 was the most stable within the range $0 \leq q \leq 1$ (Fig. 3c). The histogram of $\chi^*$ (equation (10)) showed multiple peaks.

The quantal behaviour of $\chi$ shown in Fig. 3a is understood as follows. The stall force acting on a single cargo was determined to be quantal using optical tweezers in living cells, reflecting the number of motors hauling the cargo[7,8]. From the schematic in Supplementary Fig. S2, it is understood that the drag force ($F$) becomes quantal like the stall force when the viscosity effect is large; this is thought to occur during the transport of melanosomes in melanophores. Thus, based on the assumption $F \propto \chi^*$ (equation (8)), the multiple peaks in the histogram of $\chi^*$ (Fig. 3d) were considered to reflect multiple FPUs as well as the peaks of the stall force distributions. It was then verified whether the multiple peaks indeed represent multiple FPUs in the next paragraph. As circumstantial evidence, when the number ($N_i$, where $i = 1, 2, 3$, and 4) of elements belonging to each FPU was investigated, $N_i/N_1$ (inset in Fig. 3d), whose behaviour was similar to that observed in melanosome transport in *Xenopus* melanophores[5], supported the meaning of multiple peaks of $\chi^*$ and the underlying assumption (equation (8)).

*Addition of ciliobrevin D*

To further verify whether $\chi$ can detect the number of FPUs carrying a melanosome, the number of active dyneins was intentionally decreased by adding the dynein inhibitor ciliobrevin D[25]. Figure 4a shows the time course of a melanosome in the presence of 40 μM ciliobrevin D. When we examined the time course—whose velocity was similar to that seen when the concentration of ciliobrevin ([ciliobrevin]) was 0 μM (Fig. 2b)—we found that the fluctuation ($\Delta X$) increased (inset in Fig. 4a). This change in $\Delta X$ indicated that the fluctuation reflects the effect of the decrease in dynein caused by ciliobrevin. In Fig. 4b, the velocity values of constant velocity segments were compared between cases of [ciliobrevin] = 0 and 40 μM. As was reported for peroxisome transport in *Drosophila* S2 cells[25], the high velocity portion was decreased in the presence of ciliobrevin.

The results of $\chi$ are shown for [ciliobrevin] = 10 μM (59 melanosomes from two melanophores), 20 μM (61 melanosomes from three melanophores), 30 μM (51 melanosomes from two melanophores), and 40 μM (48 melanosomes from two melanophores) (Fig. 5a–d). The number of FPUs tended to decrease with increasing [ciliobrevin]. Note that the clustering of $\chi$ was carried out by affinity propagation (Supplementary Fig. S3). The number and distribution of peaks corresponding to $\chi$ appeared to be correlated with the number of dynein motors hauling a melanosome.

We calculated the average number of FPUs from the population of each cluster (inset in Fig. 5d). The average number of FPUs decreased from 2.0 in the absence of ciliobrevin to 1.1 for [ciliobrevin] = 40 μM. It was previously reported that the velocity observed in a cytoplasmic dynein-dependent microtubule gliding assay[25] and the frequency of flagellar motility of sperm cells caused by collective movement of dynein[26] were also reduced by half for [ciliobrevin] = 40 μM. These observations of dynein-related phenomena quantitatively support non-invasive force measurement using $\chi$. When the peaks in the histogram of $\chi^*$ represent the number of FPUs, the existence of the peaks supports equation (8) because the peaks disappeared when the proportional constant (7) had a different value for each run of a melanosome.

**Discussion**

We observed the aggregation of melanosomes transported by dynein in zebrafish melanophores by bright-field microscopy at a high recording rate (800 fps). This high rate was needed to accurately measure fluctuations in the position of melanosomes. The motion of the melanosomes was quantified as fluctuation unit $\chi$ (equation (4))[11-13], which was determined on the basis of the fluctuation theorem of non-equilibrium statistical mechanics[17,21,22]. The values of $\chi$ calculated for 62 different melanosomes revealed their quantal behaviour (Fig. 3a). When equation (8) holds, the quantal behaviour of $\chi$ indicates the existence of multiple FPUs. Because force is generated by dynein, the existence of several FPUs indicates that a single melanosome is transported by multiple dynein motors. When the number of active dynein molecules was decreased by treatment with ciliobrevin D[25], the number of FPUs estimated by $\chi$ also decreased. Indeed, $\chi$ was correlated with the number of motors, supporting equations (7) and (8). Thus, $\chi$ qualifies as an indicator of force generated by motors based on its behaviour in our experiments.

While $\chi$ was found to be related to the number of FPUs (Fig. 5), the components of one FPU are still unknown. A recent cryo-electron microscopy study revealed that two dimers of dynein were connected (i.e. four monomers) via adaptor proteins, suggesting collective force generation by dynein[34]. Although the force generated by a single dynein molecule is small[9,10,30], there is a possibility that dynein molecules collectively exert a large force similar to that generated by the kinesin dimer. A future study will be necessary to elucidate the number of dynein monomers that constitute one FPU in the case of melanosome transport.

It was previously shown that velocity distributions of melanosomes in *Xenopus* melanophores had several peaks, reflecting multiple motors[5]. The velocity $v$ becomes quantal when the force $F$ of motors is quantal, and the friction coefficient $\Gamma$ is almost the same for all melanosomes through the relation $F = \Gamma v$. However, the velocity distributions measured in our experiments did not show multiple distinct peaks (Supplementary Fig. S4) because $\Gamma$ appeared to be distributed over a wide range. We speculated that $\Gamma$ had a spatial dependence in our experiment because the velocity of melanosomes increased near the nucleus where the cell thickness increased. Because we used primary cultures of melanophores from different zebrafishes instead of a melanophore cell line as in a previous study[5], and the melanophores in each experiment differed genetically, we observed variability in the viscosity of cytosol, which was influenced by melanophore structure. Additionally, in the previous study[5], actin filaments were depolymerised with latrunculin B, which may have contributed to the constant value of $\Gamma$ in the previous study.

In a previous study on melanosomes in *Xenopus* melanophores[5], it was presumed that melanosome velocity doubled with the number of FPUs. To verify this assumption in our system, we investigated changes in velocity over a long time course (Supplementary Fig. S5). Based on the measurement of $\chi$, we found that the velocity change was accompanied by a change in the number of FPUs, as expected from the previous study. On the other hand, for some melanosomes the velocity changed, while the number of FPUs did not (Supplementary Fig. S5), possibly because of a change in friction coefficient $\Gamma$ during the runs rather than a change in the number of FPUs; $\Gamma$ may exhibit spatial dependence in our case since melanosomes near the cell centre often had a higher velocity.

The noise property of *in vivo* cargo transport is another important feature of the cellular environment. In a previous study on melanosome transport[35], it was reported that the microenvironment around melanosomes is viscoelastic, based on the power spectrum densities of positions of moving melanosomes, which did not show Brownian noise behaviour. In this study, we focused only on the constant velocity segments (~ 1 s) of melanosomes (Fig. 2b and Fig. 3a) and found that the power spectrum densities for the segments showed Brownian noise behaviour for small frequencies (Supplementary Fig. S1). This Brownian noise property of $X$ in constant velocity segments was also observed for mitochondrial transport[16] and endosome transport[13] but was found to not hold when the trajectory of a cargo included pauses and bidirectional motion[16]. As a result of the Gaussian noise property in equation (3) supported by the power spectrum densities for the segments, the quantitative analysis of force could be possible. Focusing on a constant velocity segment of *in vivo* cargo transport appears to be important for quantitative analysis of its dynamics as well as qualitative analysis of subdiffusive properties of long trajectories which include the characteristic complex motion of cargo in cells.

Determination of the proportionality constant ($k_B T_{eff}$) between $\chi^*$ and $F$ is critical to estimate the *in vivo* force value produced by dynein motors as 1 FPU. In this study, we failed to determine the constant. In the case of axonal anterograde cargo transport by kinesin which was studied previously[13], the proportionality constant was estimated by comparing the $\chi^*$-$v$ relation with the known force-velocity relation of kinesin suggested by the *in vitro* stall force experiment[36]. The force-velocity relation could be fitted to the $\chi^*$-$v$ relation of the neuronal cargo by a single value of $k_B T_{eff}$[13]. Unlike the case of anterograde transport, because the value of stall force produced by dynein motors estimated in the single-molecule experiments is controversial, with reports of 1 pN[9,10,30] to 5–10 pN[31-33], the calibration of $F$ from the $\chi^*$-$v$ relation (Supplementary Fig. 6) was difficult in this study. Thus, it is essential to investigate equations (7) and (8) in combination with *in vitro* and *in vivo* stall force experiments using optical tweezers in future studies. The force-velocity relation of multiple dynein motors may lead to understanding of why $\chi^*$ takes the discrete values of 0.5, 2, 4, and 6 for 1, 2, 3, and 4 FPUs (Fig. 4a), which are different from the discrete values of $\chi^*$ in the case of axonal retrograde transport (0.1, 0.2, and 0.3)[13]. The different structures between the narrow axon and the body of eukaryotic cells, which may have wider spaces for easy cargo movement during transport, may affect the values of $\chi^*$.

**Methods**

*Culture*

Zebrafish (AB strain) were provided by and maintained at RIKEN. Cell culture and microscopic observation were performed at Tohoku University. Scales were removed from a zebrafish anaesthetised with 0.02% tolycaine and incubated for 30 min at 30°C in fish Ringer solution (5.0 mM HEPES, 116 mM NaCl, 2.9 mM KCl, 1.8 mM $CaCl_2$ [pH 7.2]) with 0.2% collagenase. The melanophores on the scales were gently removed using tweezers and transferred to a glass-bottom dish coated with Matrigel (BD Biosciences, Franklin Lakes, NJ, USA), which was filled with L15 medium (Thermo Fisher Scientific, Waltham, MA, USA) containing 1% foetal bovine serum and 1% penicillin–streptomycin (Thermo Fisher Scientific). The melanophores were cultured at 27°C. All experiments with zebrafish were conducted in compliance with the protocol approved by the Institutional Animal Care and Use Committees of Tohoku University and RIKEN.

*Addition of ciliobrevin D*

Ciliobrevin D (Calbiochem, San Diego, CA, USA) was added to the culture dish for 30 min. The melanophores were then washed twice with L15 medium, and epinephrine (final concentration: 10 μM) was added to induce melanosome aggregation.

*Melanophore observation by microscopy*

After 1 day of culture, melanophores were observed by bright-field microscopy at room temperature. Prior to observation, the melanophores were cultured in L15 medium containing 10 μM α-melanocyte-stimulating hormone for 30 min, then dispersed by washing with L15 medium without α-MSH. After adding epinephrine (final concentration: 10 μM) to the dish, images were acquired using a 100× objective lens and a sCMOS camera at 800 fps. To observe the motion of melanosomes over a wide range, a 0.5× lens was used.

      The centre position ($X$) of each melanosome was determined from the recorded images using ImageJ software (National Institutes of Health, Bethesda, MD, USA)[37]. We focused on the displacement along the direction of melanosome motion (Fig. 2a). Data were collected from 62 melanosomes from two individual melanophores in the absence of ciliobrevin; 59 melanosomes from two melanophores with 10 μM ciliobrevin; 61 melanosomes from three melanophores with 20 μM ciliobrevin; 51 melanosomes from two melanophores with 30 μM ciliobrevin; and 48 melanosomes from two melanophores with 40 μM ciliobrevin. The accuracy of the position measurement was verified using 300-nm latex beads. The mean ±

standard deviation of the position of the beads securely attached to the glass surface was $8.3 \pm 1.2$ nm (n = 4 beads).

*Theoretical background*

Microscopically, the motion of a cargo (*x*: position) at a constant velocity, for example, is described by the model

$$\gamma \frac{dx}{dt} = -\frac{\partial U_{\text{other}}(x,t)}{\partial x} - \frac{\partial U_{\text{m}}(x,t)}{\partial x} + \sqrt{2\gamma k_{\text{B}}T}\xi(t) \tag{1}$$

where $\sqrt{2\gamma k_{\text{B}}T}$ is thermal noise acting on the cargo ($k_{\text{B}}$: Boltzmann constant, $T$: temperature of the environment), $\xi$ is Gaussian noise with $\langle \xi(t)\xi(t')\rangle = \delta(t - t')$, where $<>$ denotes the time average over the time course, $U_{\text{m}}$ is an interaction between cargo and motors, and $U_{\text{other}}$ is the other interactions acting on the cargo. As assumed in the Stokes formula, $\gamma = 6\pi\eta r$ (*r*: radius of the cargo, $\eta$: viscosity of the cytosol). Note that when the environment of a cargo is viscoelastic, it is suggested that a viscoelastic memory for the friction term is added to equation (1)[38-41], instead of $U_{\text{other}}$:

$$\gamma \frac{dx}{dt} = -\frac{\partial U_{\text{m}}(x,t)}{\partial x} + \sqrt{2\gamma k_{\text{B}}T}\xi(t) - \int_{-\infty}^{t} \eta_{\text{m}}(t - t')\frac{dx(t')}{dt'}dt' + \xi_{\text{m}}(t) \tag{2}$$

where $\eta_{\text{m}}(t) \propto t^{\alpha}$ and $\langle \xi_{\text{m}}(t)\xi_{\text{m}}(t')\rangle = \eta_{\text{m}}(|t - t'|)$.

 The coarse-grained model of equation (1), which represents the motion of the cargo (*X*) for the time scale long enough to be compared with the relaxation time (~20 ms), is phenomenologically drived from equation (1) based on a previous report[42]:

$$\Gamma \frac{dX}{dt} = F + \sqrt{2\Gamma k_{\text{B}}T_{\text{eff}}}\xi(t) \tag{3}$$

The microscopic effect of $U_{\text{m}}$ and $U_{\text{other}}$ contributes to the effective viscosity $\Gamma$, effective noise $\sqrt{2\Gamma k_{\text{B}}T_{\text{eff}}}$, and the drag force $F$. The statistical properties of Gaussian force noise in equation (3) were determined from the power spectrum densities of $X$ for a small $f$ (the frequency) (Supplementary Fig. S1) and the Gaussian distribution of $P(\Delta X)$, where $\Delta X = X(t + \Delta t) - X(t)$. Subsequently, the fluctuation unit $\chi$ was considered for the macroscopic model (equation (3)).

*Calculation of $\chi$ based on the fluctuation theorem*

The fluctuation unit $\chi$, which was introduced in our previous studies[11-13], is defined as

$$\chi = \ln[P(\Delta X)/P(-\Delta X)]/\Delta X \tag{4}$$

from the distribution $P(\Delta X)$ of the displacement $\Delta X = X(t + \Delta t) - X(t)$ (e.g. Fig. 2b, inset):

$$P(\Delta X) = \exp(-(\Delta X - b)^2/2a)/(2\pi a)^{0.5} \tag{5}$$

where the fitting parameters $a$ and $b$ correspond to the variance and mean of the distribution, respectively (e.g. Fig. 2c). Note that the Brownian noise power spectrum densities of $X$ (Supplementary Fig. S1) in constant velocity segments support the use of the Gaussian function (equation (5)) to fit $P(\Delta X)$. By substituting equation (5) into equation (4),

$$\chi = 2b/a \qquad (6)$$

Thus, $\chi$ was calculated as $2b/a$ for each $P(\Delta X)$ for various intervals $\Delta t$ from 1.25 to 37.5 ms (e.g. Fig. 2d). The relaxation time of $\chi$ in Fig. 2d represents the time scale by which the motion of $X$ is described with the coarse-grained model (equation (3)). The microenvironment around the vesicle, especially its viscoelastic nature, affects the relaxation time as well as the enzymatic cycle time. The converged value ($\chi^* = \chi$ at $\Delta t = 36.25$ ms) (Fig. 2d) of $\chi$ was related to the drag force ($F$) acting on a cargo according to the equation

$$F = k_B T_{eff} \chi^* \qquad (7)$$

where $k_B$ is the Boltzmann constant, and $T_{eff}$ is the effective temperature, which is a generalised temperature in a non-equilibrium system[43-45]. A prior study[46] is helpful for derivation of the functional form of equation (7) from equation (3). Previous experiments[13] suggested that

$$F \propto \chi* \qquad (8)$$

Note that equation (6) can be interpreted as $\chi = v/D$, where $v$ and $D$ are the velocity and diffusion coefficient of a cargo. Then, $\chi$ is considered the inverse of the random parameter introduced in the reference[47].

*Smoothing and affinity propagation*

A smoothing filter was applied to the values of $\chi$ to reduce variation in the raw data for $\chi$ as a function of $\Delta t$ (Fig. 2d). We used the averaging filter

$$\chi^f(\Delta t) = (\chi(\Delta t - 1.25 \text{ ms}) + \chi(\Delta t) + \chi(\Delta t + 1.25 \text{ ms}))/3, \quad (9)$$

which is one of the simplest filters. For $\Delta t = 2.5$ ms, $\chi^f(\Delta t) = (\chi(\Delta t) + \chi(\Delta t + 1.25 \text{ ms}))/2$ was used. In this study, the converged value $\chi^*$ was defined as

$$\chi^* = \chi_f (\Delta t = 36.25 \text{ ms}). \qquad (10)$$

The data for $\chi$ ([ciliobrevin] = 0 μM) before applying the filter are shown in Supplementary Fig. S7.

Affinity propagation[48,49], an exemplar-based clustering method that does not require the number of clusters, was then adopted to cluster the smoothing-filtered two-dimensional data ($\chi^*$, $\chi_m$), where $\chi_m$ is the mean value of $\chi$ from $\Delta t = 1.25$–36.25 ms. The method was applied using the 'APCluster' function of R software[49]. The clustering was stable for the wide range of values for the parameter ($q$) (Fig. 3c).

*Power spectrum*

The power spectrum ($S(f)$) of the position $X$ of a melanosome moving at a constant velocity was calculated as

$$S(f) = \frac{\langle |X_f|^2 \rangle}{\tau_s} \tag{11}$$

$$X_f = \int_{-\tau_s/2}^{\tau_s/2} X(t) e^{i2\pi f t} dt , \tag{12}$$

where $f$ is the frequency, and $\tau s = N_w/800$ s; 800 fps is the recording rate for the pause motion, and $N_w$ (=256) is the window size (plotted in Supplementary Fig. S1). $S(f)$ for small $f$ showed Brownian noise behaviour as $S(f) \propto f^{-2}$.

*Data availability*

Data supporting the findings of this study are available within the article and its Supplementary Information files and from the corresponding author on reasonable request.

**Acknowledgements**

We thank Prof. K. Sasaki for the comments on this study. This work was supported by a grant from the Japan Agency for Medical Research and Development (no. JP17gm5810009) and Grants-in-Aid for Scientific Research (KAKENHI) from the Ministry of Education, Culture, Sports, Science, and Technology (nos. 26104501, 26115702, 26310204, and 16H00819).

**Author contributions**

K.H. designed the experiments with the assistance of Y.O. and wrote the paper. S.H. and K.H. performed the experiments and data analysis with the assistance of T.S. K.I. provided zebrafish and advice on cell culture.

**Competing interests:** The authors declare no competing interests.

**Correspondence:** Correspondence and material requests should be addressed to K.H.

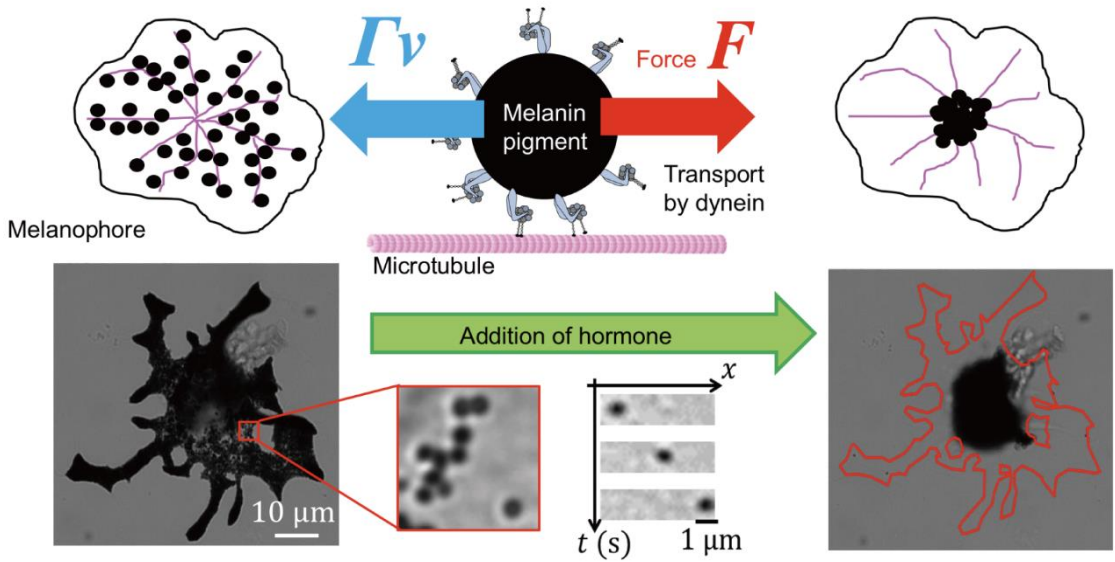

**Fig. 1** Schematic illustration of melanosome transport in a melanophore. A melanosome (black circle) is transported by multiple dynein motors (grey) along a microtubule (pink) in a melanophore. During the run at a velocity $v$, the force ($F$) generated by motors is equal to the drag force ($\Gamma v$), where $\Gamma$ is the friction coefficient of the melanosome. Addition of hormone (epinephrine) caused melanosomes to aggregate in the centre of the cell (bottom). Individual melanosomes could be tracked under a microscope. The schematic was created by the authors.

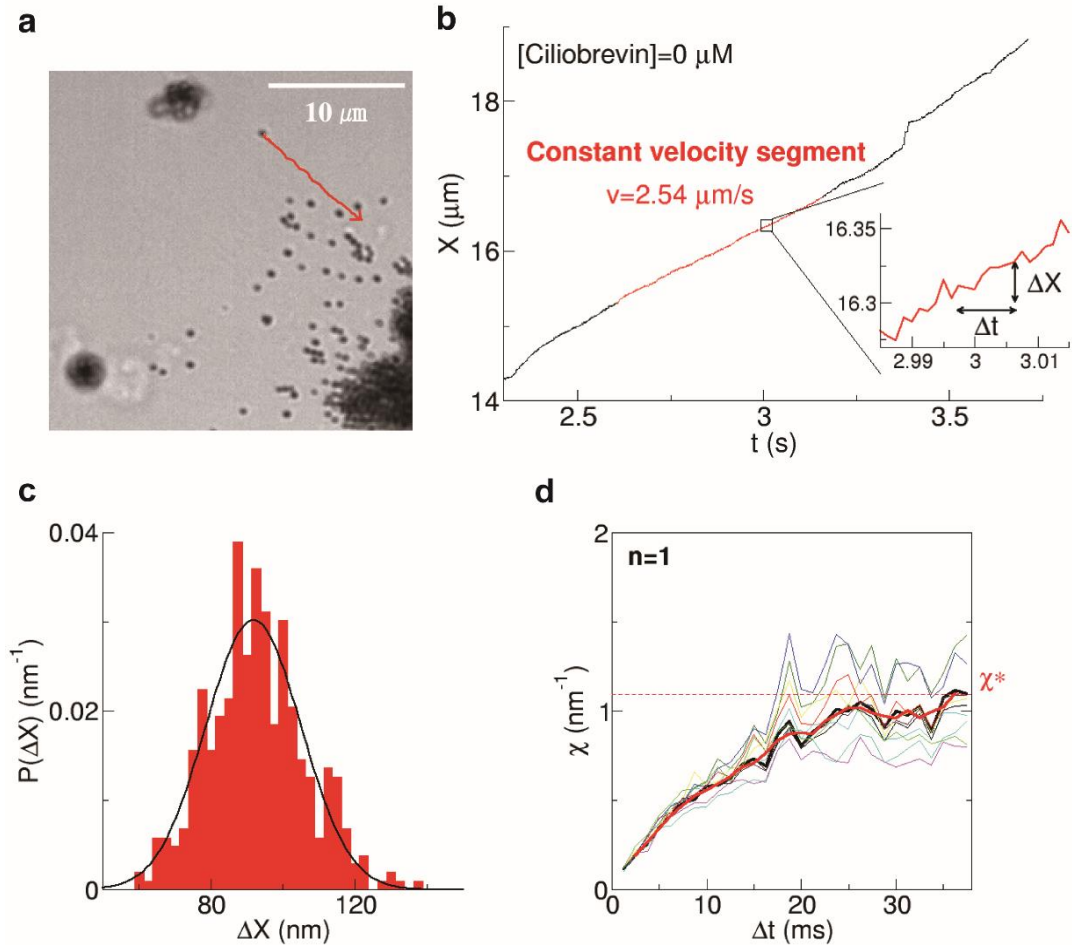

**Fig. 2** Sample calculation of a fluctuation unit ($\chi$) (equation (4)). **a** Typical motion of a melanosome after adding hormone (epinephrine). The direction of motion (red) is the positive $X$ direction. **b** Time course of the shift in the centre position ($X$) of the melanosome. When the recording rate was sufficiently high (800 fps), melanosomes were observed to fluctuate while directional motion was driven by dynein motors (inset in Fig. 2b). We focused on the constant velocity segment (red area). **c** Probability distribution ($P(\Delta X)$) of $\Delta X$ in the case of $\Delta t = 37.5$ ms, where $\Delta X = X(t + \Delta t) - X(t)$ (inset in Fig. 2b) was calculated for the constant velocity segment. $P(\Delta X)$ was fitted with a Gaussian function (black curve). **d** $\chi$ (equation (4)) calculated using the fitting parameters ($a$ and $b$) of the Gaussian function (equation (5)) plotted as a function of $\Delta t$ (thick black curve). After relaxation time, $\chi$ reaches a constant value ($\chi^*$ (equation (10)). The thick red curve represents $\chi$ after applying a smoothing filter (equation (9)). The thin curves (n = 10) represent $\chi$ calculated from different partial segments cut from the original constant velocity segments and were used to estimate the error of $\chi$ (15%).

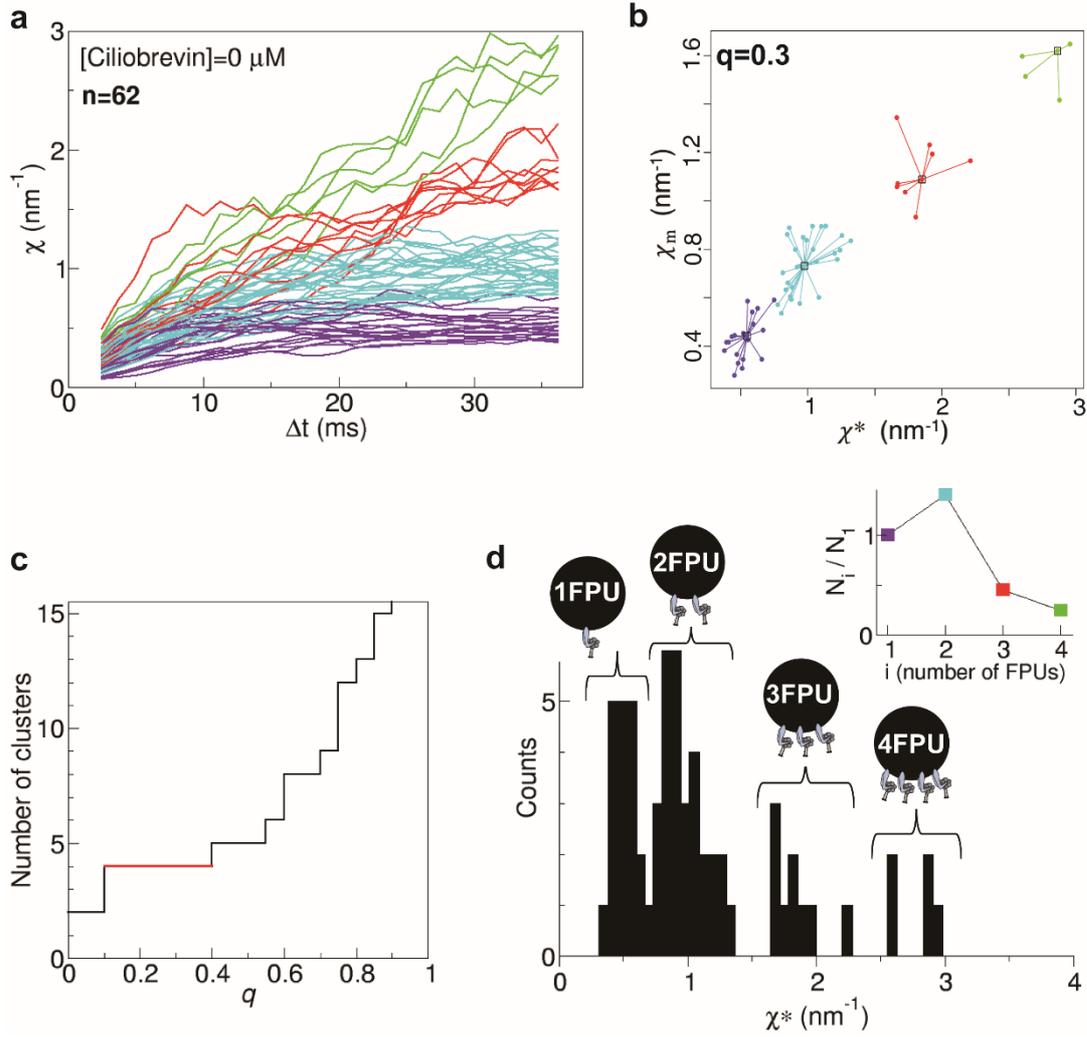

**Fig. 3** Fluctuation units ($\chi$) (equation (4)) of many melanosomes in the case of [ciliobrevin] = 0 μM. **a** The calculations in Fig. 2b–d for a single melanosome were repeated for 62 different melanosomes from two different melanophore preparations, and $\chi$ was obtained as a function of $\Delta t$ for 62 melanosomes. Each colour represents a group classified by the clustering method depicted in Fig. 3b. **b** Cluster analysis by affinity propagation in the case $q = 0.3$, where $q$ is the sole parameter of the affinity propagation analysis. $\chi$ as a function of $\Delta t$ had four clusters in the case $q = 0.3$. **c** Number of clusters obtained by affinity propagation as a function of $q$. Cluster number 4 was the most stable. **d** Histogram of $\chi^*$ (equation (10)) obtained from the values of $\chi$ in Fig. 3a. There are four FPUs for melanosome transport. (inset) The population ($N_i$, where $i = 1, 2, 3, 4$) of each FPU was calculated.

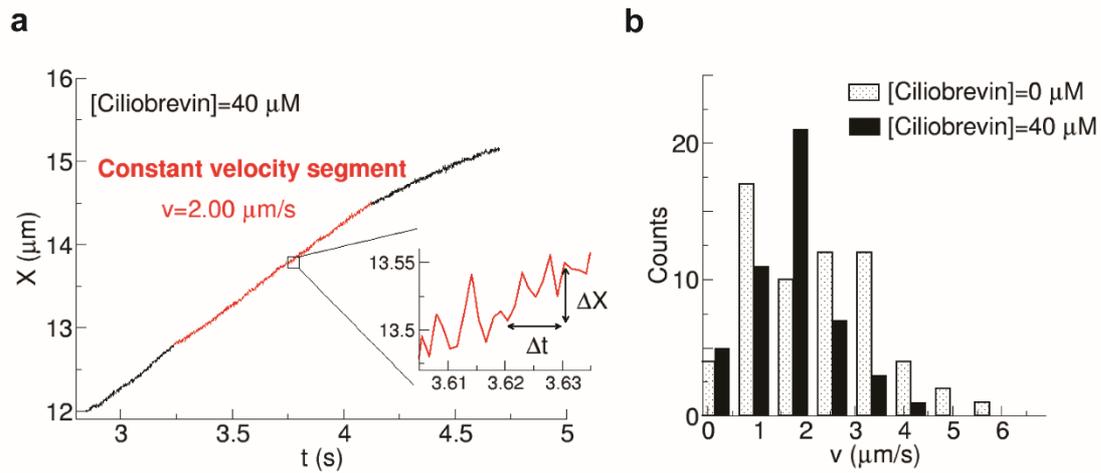

**Fig. 4** Addition of ciliobrevin D. **a** Representative time course of the change in centre position (*X*) of a melanosome in the case [ciliobrevin] = 40 μM. When the velocity was similar to that in the absence of ciliobrevin (Fig. 2b), the fluctuation in *X* was mostly increased in the case of [ciliobrevin] = 40 μM (inset). **b** Distribution of the velocity of the constant velocity segment for the cases of [ciliobrevin] = 0 (grey) and 40 μM (black).

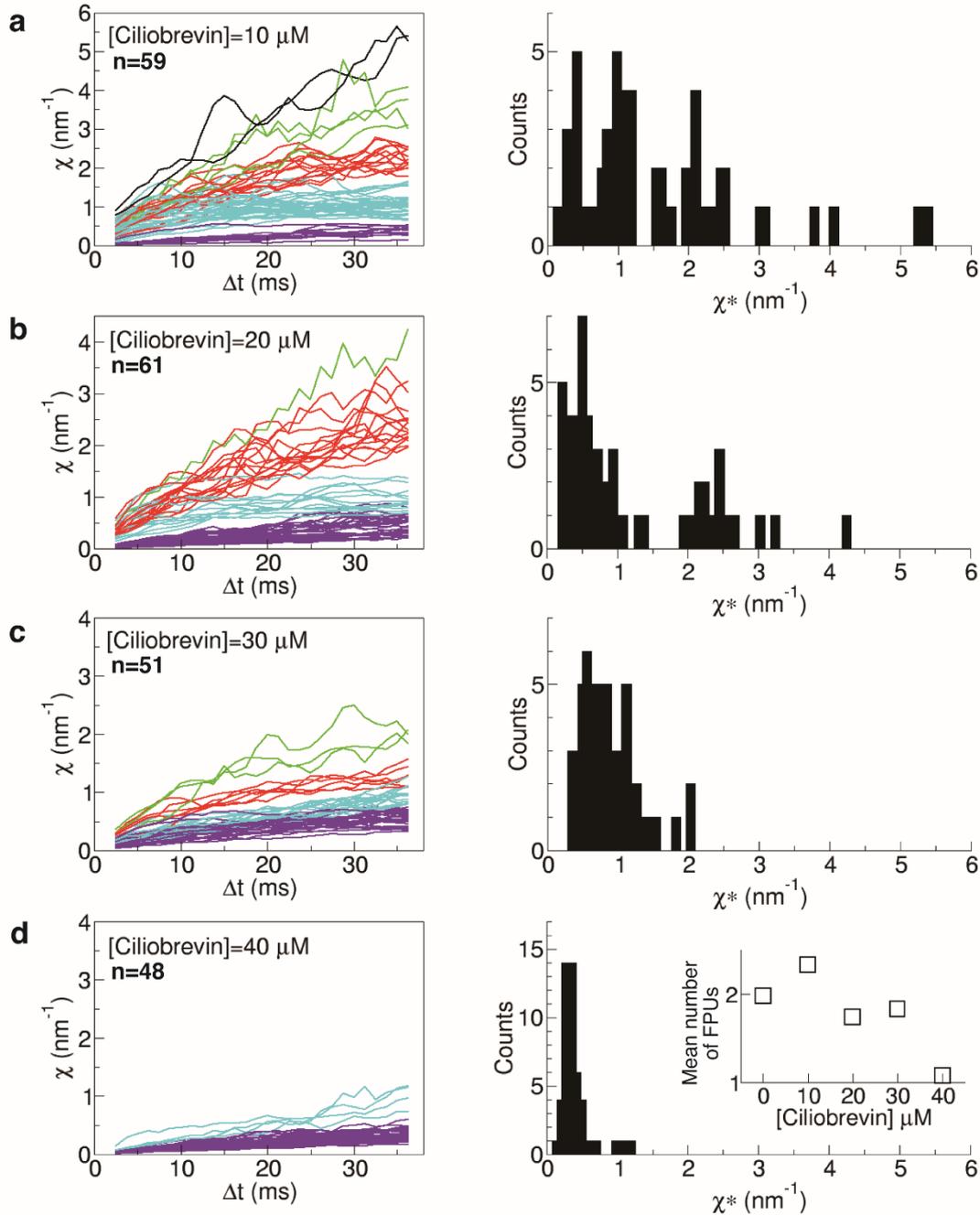

**Fig. 5** Fluctuation unit ($\chi$) (equation (4)) in the cases of [ciliobrevin] = 10 µM (59 melanosomes from two melanophores) (**a**), 20 µM (61 melanosomes from three melanophores) (**b**), 30 µM (51 melanosomes from two melanophores) (**c**), and 40 µM (48 melanosomes from two melanophores) (**d**). Left panels show $\chi$ as a function of $\Delta t$. Each colour represents a group classified by affinity propagation (Supplementary Fig. S3). Right panels show histograms of $\chi*$ (equation (10)). (inset of Fig. 5d) Mean number of FPUs calculated from the results as a function of [ciliobrevin].

**Supplementary Information**

Supplementary Figures S1–S7

**Investigation of multiple-dynein transport of melanosomes by non-invasive force measurement using fluctuation unit $\chi$**

Shin Hasegawa, Takashi Sagawa, Kazuho Ikeda, Yasushi Okada, and Kumiko Hayashi

**Supplementary Figure S1**

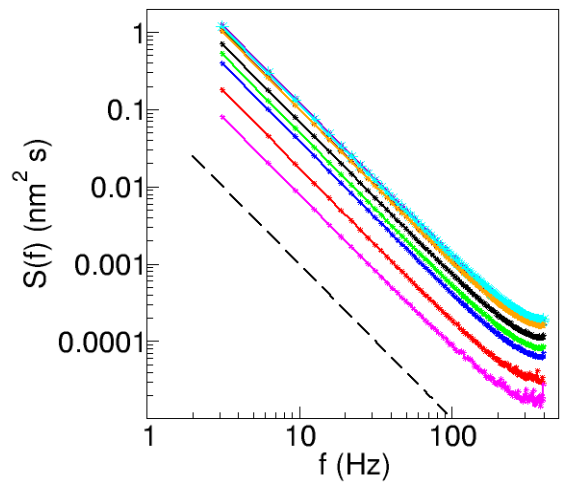

The power spectrum density (equation (11) of the main text) of the position of *X* of a melanosome for a constant velocity segment (n=10). The dotted line represents $\propto f^{-2}$.

**Supplementary Figure S2**

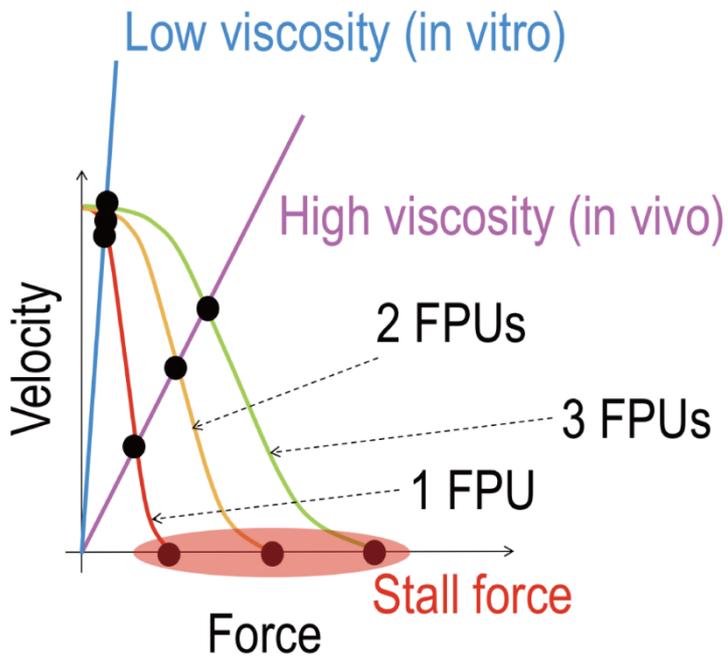

Force–velocity model of the motors considered in this study. The model was introduced in the previous study[1]. The blue and purple lines represent the relation $F = \Gamma v$ in the cases of low and high viscosities. When $\Gamma$ is large, the drag forces are considered to show a quantal behavior as well as the stall forces of the motors.

**Supplementary Figure S3**

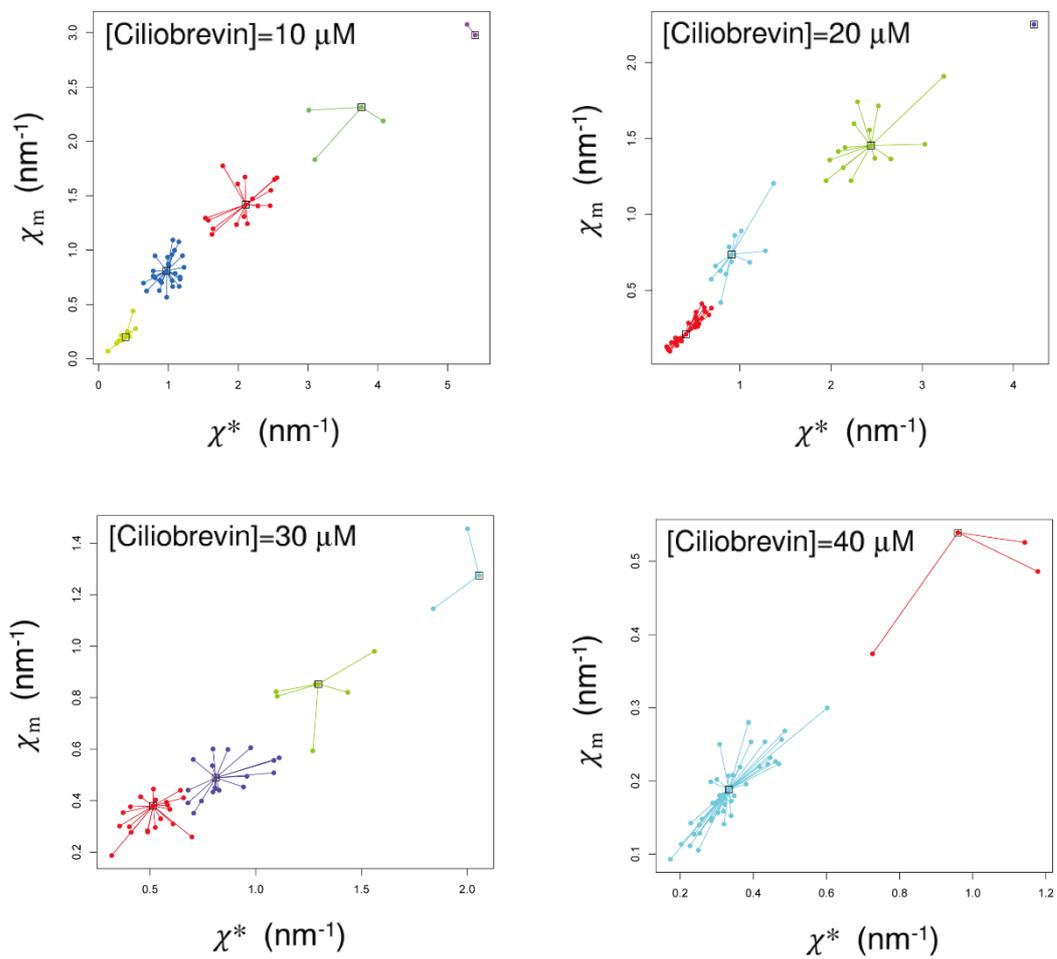

Results of the affinity propagation analysis (see Methods) with $q = 0.3$ for the data in Fig. 5. The clustering of each datum point in Fig. 5 was decided by this analysis.

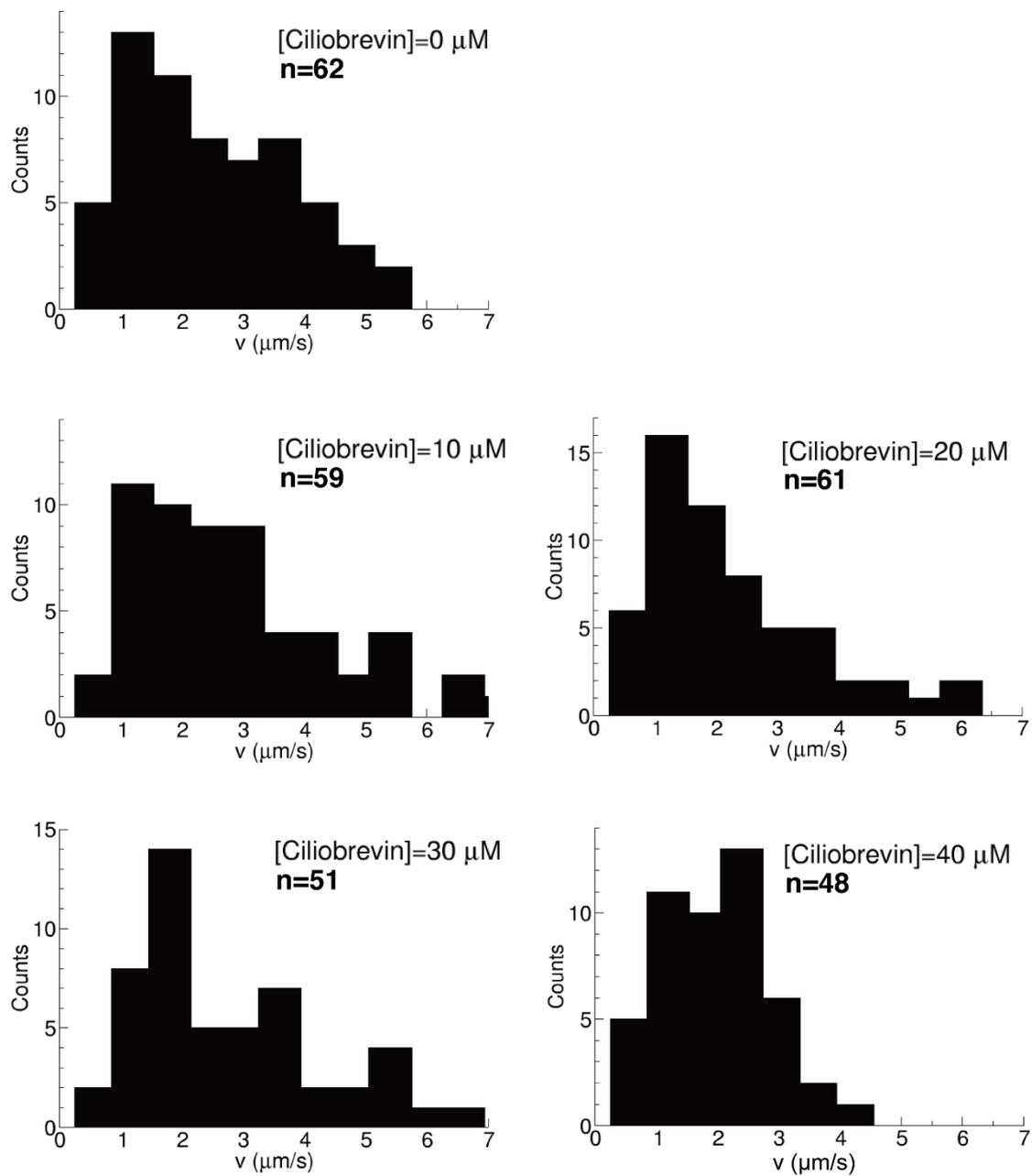

Distributions of the velocities at the constant velocity segment. The velocity distributions did not show multiple peaks clearly, unlike the drag force distributions (Fig. 4d and Fig. 5a–d).

**Supplementary Figure S5**

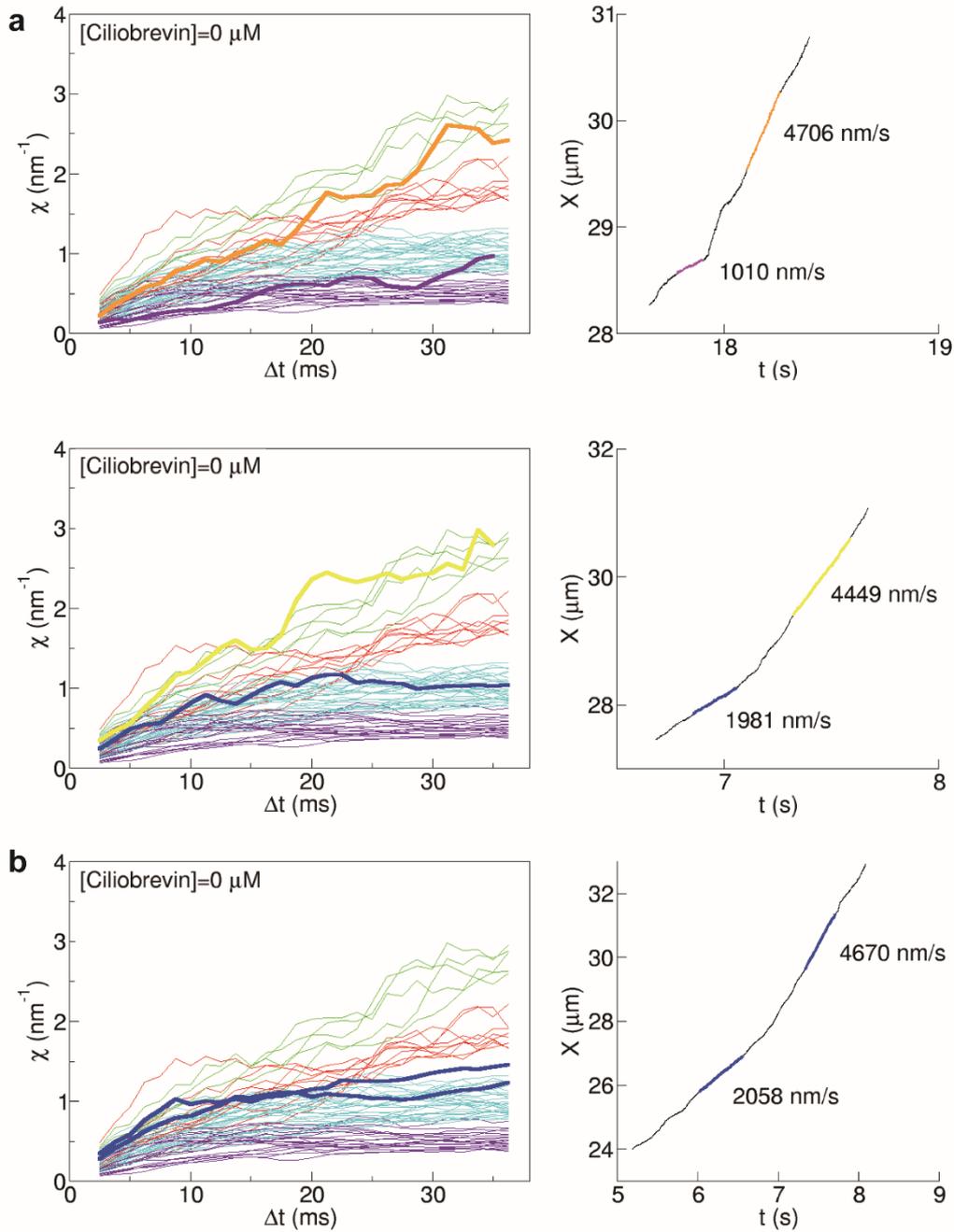

Velocity changes of the melanosomes during the aggregation process. Sometimes, velocity changes of the melanosomes were observed during the aggregation process (right panels). From the calculation of $\chi$ (left panels), we found two patterns. Here, the thick curves in the $\chi$–$\Delta t$ graphs represent $\chi$ calculated from the constant velocity segments (colored parts in the trajectories in the right-hand panels). In one case, the number of FPUs changed correspondingly to the change in velocity (**a**). In the other case, the number of FPUs did not change with the change in velocity (**b**).

**Supplementary Figure S6**

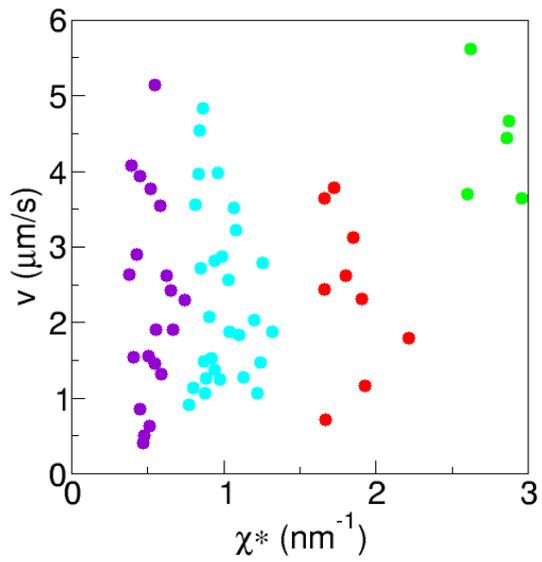

The relation between $\chi*$ and $v$ of the constant velocity segments in the case [ciliobrevin] = 0 μM. $v$ as a function of $\chi*$ investigated in Fig. 3a was plotted again for each FPU. Each color represents a cluster of Fig. 3a.

**Supplementary Figure S7**

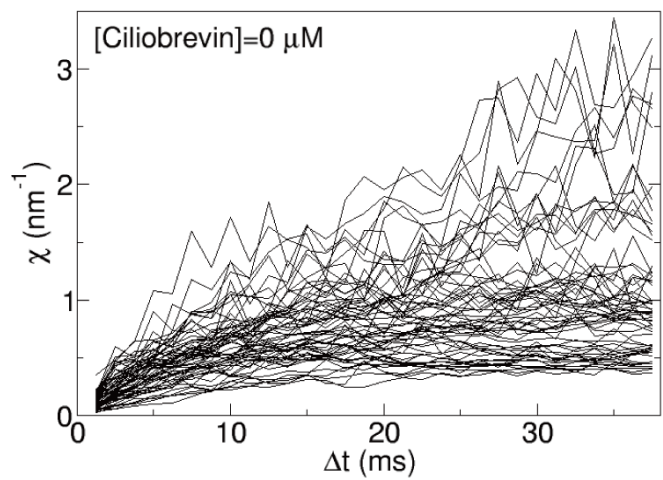

$\chi$ as a function of $\Delta t$ without the smoothing filter (see Methods). Note that the graphs in Fig. 3a represent $\chi$ after the filter.

**Supplementary References**